\documentclass[11pt,a4paper]{article}
\usepackage{ae,aecompl}
\usepackage[LGR,T1]{fontenc}
\usepackage[latin9]{inputenc}
\usepackage{textcomp}
\usepackage{mathrsfs}
\usepackage{amsmath}
\usepackage{amssymb}
\usepackage{esint}

\makeatletter

\pdfoutput=1 

\usepackage{jheppub}

\usepackage{etoolbox}
    
\patchcmd{\maketitle}{\@fpheader}{}{}{}

\usepackage{aecompl}
\usepackage{amsfonts}
\usepackage{bbm}

\title{\boldmath Mapping relativistic to ultra/non-relativistic conformal symmetries in 2D and finite $\sqrt{T\bar{T}}$ deformations}

\author[a,b]{Pablo Rodr\'{i}guez,}
\author[c]{David Tempo,}
\author[a]{Ricardo Troncoso}

\affiliation[a]{Centro de Estudios Cient\'{i}ficos (CECs), Av. Arturo Prat 514, Valdivia, Chile.}
\affiliation[b]{Departamento de Ciencias Exactas, Universidad de Los Lagos, Av. Fuchslocher 1305, Osorno, Chile.}
\affiliation[c]{Departamento  de  Ciencias  Matem\'{a}ticas  y  F\'{i}sicas, Universidad  Cat\'{o}lica  de  Temuco,  Montt  56,  Casilla  15-D,  Temuco,  Chile.}
\emailAdd{pablo.rodriguez@ulagos.cl}
\emailAdd{jtempo@uct.cl}
\emailAdd{troncoso@cecs.cl}
\preprint{CECS-PHY-20/03}

\abstract{The conformal symmetry algebra in 2D (Diff($S^{1}$)$\oplus$Diff($S^{1}$))
is shown to be related to its ultra/non-relativistic version (BMS$_{3}$$\approx$GCA$_{2}$)
through a nonlinear map of the generators, without any sort of limiting
process. For a generic classical CFT$_{2}$, the BMS$_{3}$
generators then emerge as composites built out from the chiral (holomorphic)
components of the stress-energy tensor, $T$ and $\bar{T}$, closing
in the Poisson brackets at equal time slices. Nevertheless, supertranslation
generators do not span Noetherian symmetries.
BMS$_{3}$ becomes a bona fide symmetry once the CFT$_{2}$ is marginally
deformed by the addition of a $\sqrt{T\bar{T}}$ term to the Hamiltonian.
The generic deformed theory is manifestly invariant under diffeomorphisms
and local scalings, but it is no longer a CFT$_{2}$ because its energy
and momentum densities fulfill the BMS$_{3}$ algebra. The deformation can also be described through the original
CFT$_{2}$ on a curved metric whose Beltrami differentials are determined
by the variation of the deformed Hamiltonian with respect to $T$
and $\bar{T}$. BMS$_{3}$ symmetries then arise from \textit{deformed
conformal Killing equations}, corresponding to diffeomorphisms that
preserve the deformed metric and stress-energy tensor up to local scalings.
As an example, we briefly address the deformation of $\mathrm{N}$
free bosons, which coincides with ultra-relativistic limits only for
$\mathrm{N}=1$. Furthermore, Cardy formula and
the S-modular transformation of the torus become mapped to their corresponding
BMS$_{3}$ (or flat) versions.}

\makeatother

\begin{document}
\maketitle \flushbottom \newpage{}

\section{Introduction}

Conformal symmetries enhance those of special relativity, and become
pivotal in the description of generic relativistic systems enjoying
scale invariance; see e.g., \cite{DiFrancesco:1997nk,Blumenhagen:2009zz}.
Conformal field theories (CFT's), built in terms of these extended
symmetries, are well-known to play a fundamental role in a broad variety
of subjects \cite{Cardy:1996xt,Itzykson:1988bk,Green:2012oqa,Blumenhagen:2013fgp,Polchinski:1987dy,Nakayama:2013is}.
Their power turns out to be particularly impressive in two spacetime
dimensions, as a direct consequence of the fact that the conformal
group exceptionally becomes infinite-dimensional. The conformal algebra
in 2D is described by two copies of the Witt or centerless Virasoro
algebra, being isomorphic to two copies of the algebra of diffeomorphisms
on the circle (Diff($S^{1}$)$\oplus$Diff($S^{1}$)), spanned by
\begin{eqnarray}
\left[L_{m},L_{n}\right] & = & \left(m-n\right)L_{m+n}\;\;,\;\;\left[\bar{L}_{m},\bar{L}_{n}\right]=\left(m-n\right)\bar{L}_{m+n}\;,\label{eq:ConformalAlgebra}
\end{eqnarray}
with $\left[L_{m},\bar{L}_{n}\right]=0$ and $m,n\in\mathbb{Z}$.

Another interesting accident that occurs in 2D is that the ultra and
non-relativistic limits of the conformal algebra are isomorphic (see
e.g., \cite{Bagchi:2010zz}). Intuitively, ultra/non-relativistic
limits are such that the light cone tends to shrink towards the vertical/horizontal
axis, and so one limit can be attained from the other by swapping
the role of time and space coordinates. As an additional curiosity,
the ultra/non-relativistic algebra also becomes isomorphic to the
Bondi-Metzner-Sachs one in 3D without central extensions \cite{Ashtekar:1996cd,Barnich:2006av,Duval:2014uva}.
In other words, the so-called Galilean Conformal and Conformal Carrollian
algebras in 2D turn out to be isomorphic to the BMS$_{3}$ algebra
(GCA$_{2}$$\approx$CCA$_{2}$$\approx$BMS$_{3}$), given by the
semidirect sum of the Witt algebra and supertranslations: 
\begin{align}
\left[J_{m},J_{n}\right] & =\left(m-n\right)J_{m+n}\;\;,\;\;\left[J_{m},P_{n}\right]=\left(m-n\right)P_{m+n}\;,\label{eq:BMS3-Algebra}
\end{align}
where $\left[P_{m},P_{n}\right]=0$. The algebra \eqref{eq:BMS3-Algebra}
and its centrally-extended version appeared long ago in the context
of the tensionless limit of string theory \cite{Lizzi:1986nv,Gamboa:1989px,Isberg:1993av,Bagchi:2013bga,Casali:2016atr,Casali:2017zkz,Bagchi:2017cte,Bagchi:2019cay,Bagchi:2020fpr},
and more recently in the ``flat'' analog of Liouville theory \cite{Barnich:2012rz,Barnich:2013yka}
as well as in fluid dynamics and integrable systems in 2D \cite{Penna:2017vms,Fuentealba:2017omf,Campoleoni:2018ltl}.
It also plays a leading role for nonrelativistic and flat holography
\cite{Bagchi:2009my,Barnich:2010eb,Gonzalez:2012nv,Bagchi:2012yk,Krishnan:2013wta,Bagchi:2014iea,Bagchi:2015wna,Barnich:2015dvt,Troessaert:2015syk,Bagchi:2016geg,Bagchi:2017cpu,Afshar:2016wfy,Grumiller:2016kcp,Bagchi:2021qfe},
and it emerges from the spacetime structure near generic horizons
\cite{Hawking:2016msc,Donnay:2015abr,Afshar:2015wjm,Donnay:2016ejv,Carlip:2017xne,Penna:2017bdn,Haco:2018ske,Lust:2017gez,Chandrasekaran:2018aop,Penna:2018gfx,Haco:2019ggi,Grumiller:2019fmp,Adami:2020amw,Ferreira-Martins:2021cga,Mirzaiyan:2021cwy}.
Induced, coadjoint and unitary representations have also been developed
in \cite{Barnich:2014kra,Barnich:2015uva,Campoleoni:2015qrh,Campoleoni:2016vsh}\footnote{The algebra \eqref{eq:BMS3-Algebra} also manifests as nonlocal symmetries
of a massless Klein-Gordon field in 3D \cite{Batlle:2017llu}. Different
extensions of the BMS$_{3}$ algebra have been constructed in \cite{Barnich:2014cwa,Barnich:2015sca,Banerjee:2016nio,Lodato:2016alv,Banerjee:2017gzj,Basu:2017aqn,Fuentealba:2017fck,Poojary:2017xgn,Afshar:2013vka,Gonzalez:2013oaa,Gary:2014ppa,Matulich:2014hea,Fuentealba:2015jma,Fuentealba:2015wza,Caroca:2017onr,Caroca:2019dds,Caroca:2021bjo,Adami:2020ugu,Donnay:2020fof,Batlle:2020hia,Fuentealba:2020zkf}. }.

It is worth emphasizing that the conformal algebra in 2D \eqref{eq:ConformalAlgebra}
and the BMS$_{3}$ algebra \eqref{eq:BMS3-Algebra} are not isomorphic.
Nevertheless, the latter can be obtained from the former through suitable
In\"on\"u-Wigner contractions. Indeed, changing the basis of the conformal
algebra \eqref{eq:ConformalAlgebra} according to 
\begin{equation}
P_{m}=\ell^{-1}\left(L_{m}+\bar{L}_{-m}\right)\;\;,\; \;J_{m}=L_{m}-\bar{L}_{-m}\;,
\end{equation}
one recovers the BMS$_{3}$ algebra \eqref{eq:BMS3-Algebra} in the
limit $\ell\rightarrow\infty$ (see e.g., \cite{Barnich:2012aw,Barnich:2013yka}).
Alternatively, the following change of basis: $P_{m}=\ell\left(L_{m}-\bar{L}_{m}\right)$,
$J_{m}=L_{m}+\bar{L}_{m}$ yields the same result provided that $\ell\rightarrow0$
\cite{Bagchi:2009my,Bagchi:2009pe}. The parameter $\ell$ can then
be naturally identified with the inverse of the speed of light.

\section{Map between relativistic and ultra/non-relativistic conformal algebras
in 2D}

Intriguingly, the conformal symmetry algebra in 2D \eqref{eq:ConformalAlgebra}
can be shown to be related to its ultra/non-relativistic version \eqref{eq:BMS3-Algebra}
by means of a precise nonlinear map of the generators, without the
need of performing any sort of limiting process.

In order to explicitly see the map it is useful to work in the continuum,
so that the generators of the conformal algebra \eqref{eq:ConformalAlgebra}
can be trade by two arbitrary periodic functions defined on the circle,
according to $L_{m}=\int d\phi \bar{T}\left(\phi\right)e^{-im\phi}$, $\bar{L}_{m}=\int d\phi T\left(\phi\right)e^{im\phi}$.
Thus, the conformal algebra is equivalently expressed as 
\begin{eqnarray}
\left\{ T\left(\varphi\right),T\left(\theta\right)\right\}  & = & \left(2T\left(\varphi\right)\partial_{\varphi}+\partial_{\varphi}T\left(\varphi\right)\right)\delta\left(\varphi-\theta\right)\;,\nonumber \\
\left\{ \bar{T}\left(\varphi\right),\bar{T}\left(\theta\right)\right\}  & = & -\left(2\bar{T}\left(\varphi\right)\partial_{\varphi}+\partial_{\varphi}\bar{T}\left(\varphi\right)\right)\delta\left(\varphi-\theta\right)\;,\label{eq:Conformal-Algebra-Cont}
\end{eqnarray}
with $\left\{ T\left(\varphi\right),\bar{T}\left(\theta\right)\right\} =0$,
and $\left[\cdot,\cdot\right]=i\left\{ \cdot,\cdot\right\} $. Note
that the continuous version of the conformal algebra \eqref{eq:Conformal-Algebra-Cont}
can be naturally interpreted as a Poisson structure.

The searched for mapping is then defined as follows 
\begin{equation}
P=T+\bar{T}+2\sqrt{T\bar{T}}\;\;,\;\;J=T-\bar{T}\ ,\label{map BMS to CFT-1}
\end{equation}
so that the corresponding brackets involving $J$ and $P$ can be
readily found by virtue of the ``fundamental'' ones in \eqref{eq:Conformal-Algebra-Cont},
which exactly reproduce the continuous version of the BMS$_{3}$ algebra,
given by 
\begin{align}
\left\{ J\left(\varphi\right),J\left(\theta\right)\right\}  & =\left(2J\left(\varphi\right)\partial_{\varphi}+\partial_{\varphi}J\left(\varphi\right)\right)\delta\left(\varphi-\theta\right)\;,\nonumber \\
\left\{ J\left(\varphi\right),P\left(\theta\right)\right\}  & =\left(2P\left(\varphi\right)\partial_{\varphi}+\partial_{\varphi}P\left(\varphi\right)\right)\delta\left(\varphi-\theta\right)\;,\label{eq:BMS3-Algebra-Cont}
\end{align}
with $\left\{ P\left(\varphi\right),P\left(\theta\right)\right\} =0$.
In Fourier modes, $J_{m}=\int d\phi J\left(\phi\right)e^{im\phi}$,
$P_{m}=\int d\phi P\left(\phi\right)e^{im\phi}$, the algebra \eqref{eq:BMS3-Algebra-Cont}
then reduces to \eqref{eq:BMS3-Algebra}.

Bearing in mind that supertranslation generators are defined up to
a global scale factor, making $P\rightarrow\alpha P$ with constant
$\alpha$ in the map \eqref{map BMS to CFT-1}, yields the same result.
Thus, for simplicity and later convenience, we keep assuming $\alpha=1$
afterwards.

In sum, the nonisomorphic conformal and BMS$_{3}$ algebras, in \eqref{eq:Conformal-Algebra-Cont}
and \eqref{eq:BMS3-Algebra-Cont} respectively, are nonlinearly related
by virtue of the map defined through \eqref{map BMS to CFT-1}, and
it is worth highlighting that that no limiting process is involved
in the mapping.
\section{BMS$_{3}$ generators within CFT$_{2}$}

The mapping in \eqref{map BMS to CFT-1} naturally makes one wondering
about how precisely the BMS$_{3}$ algebra manifests itself for a
generic (nonanomalous) classical CFT$_{2}$. Indeed, the mapping directly
prescribes a way in which BMS$_{3}$ generators emerge as composites
of those of the conformal symmetries. Nonetheless, it can be shown
that the composite generators do not span a Noetherian symmetry of
the CFT$_{2}$.

In order to see that, let us consider a generic CFT$_{2}$ on a cyllinder.
In the conformal gauge, using null coordinates $x=t+\phi$ and $\bar{x}=t-\phi$,
the canonical generators of the conformal symmetries are given by
(see e.g., \cite{DiFrancesco:1997nk,Blumenhagen:2009zz})

\begin{equation}
{\cal Q}_{CFT}\left[\epsilon,\bar{\epsilon}\right]=\int d\phi\left(\epsilon T+\bar{\epsilon}\bar{T}\right)\;,\label{eq:Q-CFT2}
\end{equation}
being conserved ($\dot{\mathcal{Q}}_{CFT}=0$) by virtue of the (anti-)chirality
of the components of the stress-energy tensor and the parameters ($\partial\bar{T}=\bar{\partial}T=\partial\bar{\epsilon}=\bar{\partial}\epsilon=0$).
The transformation laws of $T$ and $\bar{T}$ then read from the
conformal algebra \eqref{eq:Conformal-Algebra-Cont}, since $\delta_{\eta_{1}}{\cal Q}_{CFT}\left[\eta_{2}\right]=\left\{ {\cal Q}_{CFT}\left[\eta_{2}\right],{\cal Q}_{CFT}\left[\eta_{1}\right]\right\} $
with $\eta_{i}=\left(\epsilon_{i},\bar{\epsilon}_{i}\right)$, so
that 
\begin{equation}
\delta T=2T\partial\epsilon+\partial T\epsilon\;\;,\;\;\delta\bar{T}=2\bar{T}\bar{\partial}\bar{\epsilon}+\bar{\partial}\bar{T}\bar{\epsilon}\ .\label{eq:deltaTyTbar}
\end{equation}

The nonlinear map \eqref{map BMS to CFT-1} implies that the generators
\eqref{eq:Q-CFT2} and transformation laws \eqref{eq:deltaTyTbar},
can be expressed as 
\begin{eqnarray}
{\cal Q}_{CFT}\left[\epsilon,\bar{\epsilon}\right] & = & \int d\phi\left(\epsilon_{J}J+\epsilon_{P}P\right)\;,\label{eq:Q-CFT2-bms3}
\end{eqnarray}
\begin{equation}
\delta P=2P\epsilon_{J}^{\prime}+P^{\prime}\epsilon_{J}\;\;,\;\;\delta J=2P\epsilon_{P}^{\prime}+P^{\prime}\epsilon_{P}+2J\epsilon_{J}^{\prime}+J^{\prime}\epsilon_{J}\ ,\label{eq:deltaPJ-BMS3}
\end{equation}
where prime stands for $\partial_{\phi}$, while the parameters $\epsilon_{J}$,
$\epsilon_{P}$, relate to $\epsilon$ and $\bar{\epsilon}$ through
\begin{equation}
\epsilon=\epsilon_{J}+\left(1+\sqrt{\frac{\bar{T}}{T}}\right)\epsilon_{P}\;\;,\;\;\bar{\epsilon}=-\epsilon_{J}+\left(1+\sqrt{\frac{T}{\bar{T}}}\right)\epsilon_{P}\;.\label{map paremeters CFT to BMS}
\end{equation}
Thus, the generators and transformation laws in \eqref{eq:Q-CFT2-bms3},
\eqref{eq:deltaPJ-BMS3}, acquire the expected form of those for the
BMS$_{3}$ algebra (see e.g., \cite{Barnich:2012rz,Fuentealba:2017omf})\footnote{A warning note is in order: if the parameters $\epsilon$ and $\bar{\epsilon}$
were still assumed to be chiral, this would be just a mirage; because
in that case, the new ones, $\epsilon_{J}$, $\epsilon_{P}$, would
become state-dependent, and hence, this would only amount to an alternative
way of expressing the original conformal algebra generators and transformation
laws in \eqref{eq:Q-CFT2}, \eqref{eq:deltaTyTbar}, in terms of different
variables.}.

Legitimate BMS$_{3}$ generators are obtained when the parameters
$\epsilon$, $\bar{\epsilon}$ are no longer chiral, but instead,
being manifestly state-dependent according to \eqref{map paremeters CFT to BMS}.
Hence, at fixed time slices, the parameters $\epsilon_{J}$, $\epsilon_{P}$
can be consistently assumed to be state-independent arbitrary functions,
so that the Poisson brackets of the generators 
\begin{eqnarray}
 & {\cal \tilde{Q}}\left[\epsilon_{J},\epsilon_{P}\right]= & \int d\phi\left(\epsilon_{J}J+\epsilon_{P}P\right)\;,\label{eq:Qbms}
\end{eqnarray}
clearly close according to the BMS$_{3}$ algebra by virtue of \eqref{eq:BMS3-Algebra-Cont}.

It is worth emphasizing that since $J$ stands for the momentum density,
superrotation generators yield the corresponding conserved charges.
Nevertheless, supertranslation generators are not conserved, as it
can bee seen from the time evolution of $P$, that can be obtained
from that of the (anti-)chiral $T$ and $\bar{T}$ by virtue of the
map \eqref{map BMS to CFT-1}, given by 
\begin{equation}
\dot{P}=2J^{\prime}-J\left(\log P\right)^{\prime}\ .
\end{equation}
Therefore, supertranslations do not correspond to Noetherian symmetries
of the CFT$_{2}$.

\section{BMS$_{3}$ symmetries from $\sqrt{T\bar{T}}$ deformations}

According to the map \eqref{map BMS to CFT-1}, the supertranslation
density $P$ can be seen as a finite nontrivial marginal deformation
of the CFT$_{2}$ energy density $H=T+\bar{T}$. Hence, a simple way
to achieve conservation of supertranslations consists in deforming
the original Hamiltonian of the CFT$_{2}$ to coincide with the supertranslation
generator. Thus, starting from the CFT$_{2}$ in the conformal gauge,
the simplest deformation is implemented through the Hamiltonian density
$\tilde{H}=H+2\sqrt{T\bar{T}}=P$, so that the deformed action reads
\begin{equation}
\tilde{I}=I_{CFT}-\int dxd\bar{x}\sqrt{T\bar{T}}\ .\label{eq:I-tilde-sqrtTTbar}
\end{equation}
Note that since only the Hamiltonian was deformed, the Poisson brackets
remain the same as those of the original CFT$_{2}$ in \eqref{eq:Conformal-Algebra-Cont}.
Hence, the time evolution of supertranslation and superrotation densities
can be readily obtained from \eqref{eq:BMS3-Algebra-Cont} 
\[
\dot{P}=\left\{ P,\tilde{\mathscr{H}}\right\} =0\;\;,\;\;\dot{J}=\left\{ J,\tilde{\mathscr{H}}\right\} =P^{\prime}\ ,
\]
with $\tilde{\mathscr{H}}=\int d\phi P$; so that the canonical BMS$_{3}$
generators \eqref{eq:Qbms} are now manifestly conserved ($\dot{\tilde{\mathcal{Q}}}=0$)
provided that the parameters fulfill $\dot{\epsilon}_{P}=\epsilon_{J}^{\prime}$
and $\dot{\epsilon}_{J}=0$, being apparently state independent.

Therefore, the BMS$_{3}$ generators \eqref{eq:Qbms} span a bona
fide Noetherian symmetry of the deformed action \eqref{eq:I-tilde-sqrtTTbar}.

It is also worth pointing out that the deformed theory \eqref{eq:I-tilde-sqrtTTbar}
retains the integrability properties of the original CFT$_{2}$, since
the universal enveloping algebra of BMS$_{3}$ also contains an infinite
number of independent commuting (KdV-like) charges \cite{Fuentealba:2017omf}\footnote{This last property resembles that of the $T\bar{T}$ deformation \cite{Zamolodchikov:2004ce,Smirnov:2016lqw,Cavaglia:2016oda}
(being widely studied in e.g., \cite{McGough:2016lol,Giribet:2017imm,Kraus:2018xrn,Bonelli:2018kik,Taylor:2018xcy,Datta:2018thy,Babaro:2018cmq,Conti:2018jho,Aharony:2018bad,Araujo:2018rho,Gorbenko:2018oov,Conti:2019dxg,Coleman:2019dvf,Guica:2019nzm,Callebaut:2019omt,Apolo:2019zai,Jorjadze:2020ili,Leoni:2020rof,Caputa:2020lpa,Babaei-Aghbolagh:2020kjg,Kraus:2021cwf});
nonetheless, some differences must be stressed. Indeed, in that case
the conformal weight of the deformation implies that it is an irrelevant
one, also depending on a single continuous parameter, and where $T$
and $\bar{T}$ stand for those of the deformed theory; while in our
case, they correspond to those of the original CFT$_{2}$ and yield
to a rigid finite marginal deformation.}.

For a generic gauge choice, the deformation \eqref{eq:I-tilde-sqrtTTbar}
can be written as 
\begin{equation}
\tilde{I}=I_{CFT}-\int d^{2}x\sqrt{\det T_{\mu\nu}}\;.\label{eq:I-tilde-detTmunu}
\end{equation}
where it is implicitly assumed that $I_{CFT}$ is written in Hamiltonian
form, and $T_{\mu\nu}$ stands for the stress-energy tensor of the
undeformed CFT$_{2}$. Remarkably, the action \eqref{eq:I-tilde-detTmunu}
keeps being invariant under diffeomorphisms and local scalings, but
it is no longer a CFT$_{2}$ because the energy and momentum densities
of the deformed theory yield to generators that fulfill the BMS$_{3}$
algebra \eqref{eq:BMS3-Algebra-Cont} instead of the conformal one
in \eqref{eq:Conformal-Algebra-Cont}.

In order to see that, let us consider the original CFT$_{2}$ in a
generic (non-conformal) gauge, so that in a local patch, the two-dimensional
metric can be brought to the same conformal class as the following
one c.f., \cite{Teitelboim:1983ux} 
\begin{equation}
ds^{2}=-N^{2}dt^{2}+\left(d\phi+N^{\phi}dt\right)^{2}\ ,\label{eq:ds2-N-Nphi}
\end{equation}
where $N$ and $N^{\phi}$ stand for the lapse and shift functions,
respectively\footnote{In null (holomorphic) coordinates, this would amount to switch on
the Beltrami differentials.}. The total Hamiltonian of the CFT$_{2}$ then reads 
\begin{align}
\mathscr{H}_{CFT}&=\int d\phi\left[N\left(T+\bar{T}\right)+N^{\phi}\left(T-\bar{T}\right)\right]=\int d\phi\left(NH+N^{\phi}J\right)\ .\label{eq:H-CFT-GenericGauge}
\end{align}
The deformation in \eqref{eq:I-tilde-detTmunu} has the net effect
of deforming the energy density of the CFT$_{2}$ to be that of a
supertranslation, i.e., $H\rightarrow P$, so that the total Hamiltonian
deforms as $\mathscr{H}_{CFT}\rightarrow\mathscr{\tilde{H}}$, with
\begin{equation}
\mathscr{\tilde{H}}=\int d\phi\left(NP+N^{\phi}J\right)\ .\label{eq:HbmsTotal}
\end{equation}
Supertranslation and superrotation densities evolution is then spanned
by the deformed Hamiltonian $\tilde{\mathscr{H}}$, which by virtue
of \eqref{eq:BMS3-Algebra-Cont} reads 
\begin{align}
\dot{P} & = \{P,\mathscr{\tilde{H}} \}=2PN^{\phi\prime}+P^{\prime}N^{\phi}\;,\nonumber \\
\dot{J} & = \{ J,\mathscr{\tilde{H}} \} =2PN^{\prime}+P^{\prime}N+2JN^{\phi\prime}+J^{\prime}N^{\phi}\ \label{eq:P-J-Punto-BMS-N-Nphi}.
\end{align}
In absence of global obstructions, the canonical generators become
expressed as an integral over the spatial circle precisely as in \eqref{eq:Qbms},
but now being conserved provided that the state-independent parameters
fulfill 
\begin{equation}
\dot{\epsilon}_{P}=N\epsilon_{J}^{\prime}-N^{\prime}\epsilon_{J}+N^{\phi}\epsilon_{P}^{\prime}-N^{\phi\prime}\epsilon_{P}\;\;,\;\;\dot{\epsilon}_{J}=N^{\phi}\epsilon_{J}^{\prime}-N^{\phi\prime}\epsilon_{J}\ .\label{eq:EpsilonBMS-Punto-N-Nphi}
\end{equation}
Thus, the transformation law of supertranslation and superrotation
densities is given by \eqref{eq:deltaPJ-BMS3}, corresponding to Noetherian
BMS$_{3}$ symmetries.

\section{Geometric aspects}

Since the deformed action is manifestly invariant under diffeomorphisms
$\xi=\xi^{\mu}\partial_{\mu}$, it is reassuring to verify that the
Noether current $j^{\mu}=\tilde{{\cal T}}_{\;\;\nu}^{\mu}\xi^{\nu}$,
with 
\begin{equation}
\tilde{{\cal T}}_{\;\;\nu}^{\mu}=\left(\begin{array}{cc}
NP+N^{\phi}J & J\\
-N^{\phi}\left(N^{\phi}J+2NP\right) & -\left(NP+N^{\phi}J\right)
\end{array}\right)\ ,\label{eq:Tmunu-density-bms}
\end{equation}
is conserved ($\partial_{\mu}j^{\mu}=0$) provided that the evolution
equations of the energy and momentum densities \eqref{eq:P-J-Punto-BMS-N-Nphi},
as well as those of the parameters in \eqref{eq:EpsilonBMS-Punto-N-Nphi}
hold. The precise form of the diffeomorphisms is then identified as
\begin{equation}
\xi^{\mu}=N^{-1}\left(\epsilon_{P},N\epsilon_{J}-N^{\phi}\epsilon_{P}\right)\ ,\label{BMS3-Vector}
\end{equation}
which close in the Lie brackets, $\left[\xi_{1},\xi_{2}\right]=\xi_{3}$,
with 
\begin{equation}
\epsilon_{P}^{3}=\epsilon_{J}^{1}\left(\epsilon_{P}^{2}\right)^{\prime}+\epsilon_{P}^{1}\left(\epsilon_{J}^{2}\right)^{\prime}-\left(1\leftrightarrow2\right)\;\;,\;\;\epsilon_{J}^{3}=\epsilon_{J}^{1}\left(\epsilon_{J}^{2}\right)^{\prime}-\left(1\leftrightarrow2\right)\ ,
\end{equation}
according to the BMS$_{3}$ algebra when the parameters $\epsilon_{P}^{i}$
, $\epsilon_{J}^{i}$ obey \eqref{eq:EpsilonBMS-Punto-N-Nphi}.

Note that one might be tempted to extract an stress-energy tensor $\Theta_{\;\;\nu}^{\mu}$
from the corresponding density in \eqref{eq:Tmunu-density-bms} by
making use of the metric of the undeformed theory $g_{\mu\nu}$ in
\eqref{eq:ds2-N-Nphi}, according to $\tilde{{\cal T}}_{\;\;\nu}^{\mu}=\sqrt{-g}\Theta_{\;\;\nu}^{\mu}$.
However, this tensor is not conserved ($\nabla_{\mu}\Theta_{\ \;\nu}^{\mu}\neq0$),
reflecting the fact that the metric the of CFT$_{2}$ is not preserved
under BMS$_{3}$ diffeomorphisms $\xi^{\mu}$ up to a local scaling,
i.e., 
\begin{equation}
\nabla_{\mu}\xi_{\nu}+\nabla_{\nu}\xi_{\mu}-\lambda g_{\mu\nu}\neq0\ .
\end{equation}
Hence, the metric of the undeformed CFT$_{2}$ is not a suitable object
to describe the geometric properties of the deformed theory.

An appropriate Riemannian metric for the geometric description of
the deformation is obtained as follows. Note that the total deformed
Hamiltonian \eqref{eq:HbmsTotal} is a homogeneous functional of $T$
and $\bar{T}$ of degree one, so that it fulfills the following identity
\begin{equation}
\tilde{\mathscr{H}}=\int d\phi\left(\frac{\delta\tilde{\mathscr{H}}}{\delta T}T+\frac{\delta\tilde{\mathscr{H}}}{\delta\bar{T}}\bar{T}\right)=\int d\phi\left(\frac{\delta\tilde{\mathscr{H}}}{\delta H}H+\frac{\delta\tilde{\mathscr{H}}}{\delta J}J\right)\ .
\end{equation}
Therefore, the deformed theory can be equivalently described by placing
the original CFT$_{2}$ on a state-dependent curved metric, whose
lapse and shift functions, $\tilde{N}$ and $\tilde{N}^{\phi}$, are
respectively given by the variation of the deformed Hamiltonian with
respect to the energy and momentum densities of the undeformed theory,
i.e.,\footnote{Beltrami differentials are
determined by the variation of the deformed Hamiltonian with respect
to $T$ and $\bar{T}$. } 
\begin{equation}
d\tilde{s}^{2}=-\left(\frac{\delta\tilde{\mathscr{H}}}{\delta H}\right)^{2}dt^{2}+\left(d\phi+\frac{\delta\tilde{\mathscr{H}}}{\delta J}dt\right)^{2}\ .\label{eq:Deformed-Metric-Generic}
\end{equation}
The mapping \eqref{map BMS to CFT-1} allows to express the deformed
metric \eqref{eq:Deformed-Metric-Generic} in terms of the supertranslation
and superrotation densities, so that it reads 
\begin{align}
d\tilde{s}^{2}=&-N^{2}\left(\frac{2P^{2}}{J^{2}-P^{2}}\right)^{2}dt^{2} +\left(d\phi+\left(N^{\phi}+N\frac{2JP}{J^{2}-P^{2}}\right)dt\right)^{2}\ ,\label{eq:deformed-metric-bms}
\end{align}
where $N$ and $N^{\phi}$ correspond to the (state-independent) lapse
and shift functions of the original undeformed metric in \eqref{eq:ds2-N-Nphi},
respectively\footnote{Note that the Ricci scalar of the deformed metric $\tilde{g}_{\mu\nu}$
differs from that of the undeformed one $g_{\mu\nu}$ ($\tilde{R}\neq R$).
In contradistinction, the corresponding metrics in the geometric interpretation
of the $T\bar{T}$ deformation \cite{Dubovsky:2017cnj,Cardy:2018sdv,Dubovsky:2018bmo,Conti:2018tca}
are related through state-dependent diffeomorphisms.}.

It must be emphasized that the manifest state dependence of the lapse
and shift functions (or Beltrami differentials) of the deformed metric
\eqref{eq:deformed-metric-bms} provides a local obstruction to gauge
them away, preventing the possibility of choosing the standard conformal
gauge once the theory is deformed.

A proper stress-energy tensor $\tilde{\Theta}_{\ \ \nu}^{\mu}$, consistent
with invariance under diffeomorphisms and local scalings of the deformed
action \eqref{eq:I-tilde-detTmunu}, is then readily obtained from
$\tilde{{\cal T}}_{\;\;\nu}^{\mu}=\sqrt{-\tilde{g}}\tilde{\Theta}_{\;\;\nu}^{\mu}$,
where $\tilde{g}_{\mu\nu}$ stands for the state-dependent metric
in \eqref{eq:deformed-metric-bms}. Indeed, the deformed stress-energy
tensor fulfills 
\begin{equation}
\tilde{\Theta}_{\mu\nu}=\tilde{\Theta}_{\nu\mu}\;\;,\;\;\tilde{\Theta}_{\;\mu}^{\mu}=0\;\;,\;\;\tilde{\nabla}_{\mu}\tilde{\Theta}_{\ \nu}^{\mu}=0\;,
\end{equation}
being automatically symmetric and traceless, while its conservation
implies the evolution equations of supertranslation and superrotation
densities \eqref{eq:P-J-Punto-BMS-N-Nphi}. Therefore, the canonical
BMS$_{3}$ generators \eqref{eq:Qbms} can be written in manifestly
covariant way as 
\begin{equation}
\tilde{{\cal Q}}\left[\epsilon_{J},\epsilon_{P}\right]=\int d\phi\sqrt{\tilde{\gamma}}\tilde{n}_{\mu}\tilde{\Theta}_{\;\;\nu}^{\mu}\xi^{\nu}\;,
\end{equation}
with $\xi^{\mu}$ given by \eqref{BMS3-Vector}, and according to
the deformed metric $\tilde{g}_{\mu\nu}$ in \eqref{eq:deformed-metric-bms},
the unit timelike normal is given by $\tilde{n}_{\mu}=(\tilde{N},0)$,
and $\tilde{\gamma}=1$.

The geometric description of the deformed theory is then suitably
carried out in terms of the two relevant structures, $\tilde{g}_{\mu\nu}$
and $\tilde{\Theta}_{\;\;\nu}^{\mu}$, being inextricably intertwined.
In fact, since both objects are state dependent, they acquire nontrivial
functional variations when acting on them under diffeomorphisms, given
by 
\begin{align}
\delta_{\xi}\tilde{g}_{\mu\nu}=\frac{\delta\tilde{g}_{\mu\nu}}{\delta P}\delta_{\xi}P+\frac{\delta\tilde{g}_{\mu\nu}}{\delta J}\delta_{\xi}J\;\;,\;\;\delta_{\xi}\tilde{\Theta}_{\mu\nu}=\frac{\delta\tilde{\Theta}_{\mu\nu}}{\delta P}\delta_{\xi}P+\frac{\delta\tilde{\Theta}_{\mu\nu}}{\delta J}\delta_{\xi}J\;.\label{eq:FunctionalVariations-g-theta-tilde}
\end{align}
Therefore, since the functional variations \eqref{eq:FunctionalVariations-g-theta-tilde}
must be taken into account, BMS$_{3}$ symmetries geometrically arise
from diffeomorphisms $\xi$ that preserve the form of both relevant
structures up to a local scaling, i.e., from the solutions of the
following \textit{deformed conformal Killing equations} 
\[
\tilde{\nabla}_{\mu}\xi_{\nu}+\tilde{\nabla}_{\nu}\xi_{\mu}-\lambda\tilde{g}_{\mu\nu}=\delta_{\xi}\tilde{g}_{\mu\nu}\ ,
\]
\begin{equation}
\mathcal{L}_{\xi}\tilde{\Theta}_{\mu\nu}=\delta_{\xi}\tilde{\Theta}_{\mu\nu}\ ,\label{Deformed-Conf. -Killing equations}
\end{equation}
where $\mathcal{L}_{\xi}$ stands for the Lie derivative.

It is amusing to verify that starting from scratch with the deformed
metric and stress-energy tensor, $\tilde{g}_{\mu\nu}$ and $\tilde{\Theta}_{\;\;\nu}^{\mu}$,
the deformed conformal Killing equations \eqref{Deformed-Conf. -Killing equations}
can be exactly solved. Indeed, the solution is precisely given by
the BMS$_{3}$ diffeomorphisms $\xi^{\mu}$ in \eqref{BMS3-Vector}
with parameters $\epsilon_{P}$, $\epsilon_{J}$ fulfilling \eqref{eq:EpsilonBMS-Punto-N-Nphi},
where the transformation law of supertranslation and superrotation
densities is also found to be given by \eqref{eq:deltaPJ-BMS3}.

Note that the geometric interpretation also allows to find the transformation
law of the fields in the deformed theory from those of the original
undeformed (primary) fields, collectively denoted by $\chi$, by writing
them in a manifestly covariant way, and then acting with the Lie derivative
along BMS$_{3}$ symmetries spanned by $\xi$, i.e., $\delta_{\xi}\chi=\mathcal{L}_{\xi}\chi$.

\section{Deformed free bosons}

Let us see how the deformation works in a simple and concrete example,
given by the action of $\mathrm{N}$ free bosons with flat target
metric,

\begin{equation}
I\left[\Phi^{I}\right]=-\frac{1}{2}\int d^{2}x\sqrt{-g}\delta_{IK}\partial_{\mu}\Phi^{I}\partial^{\mu}\Phi^{K}\ .\label{eq:I-NfreeBs-CFT}
\end{equation}
Before implementing the generic deformation \eqref{eq:I-tilde-detTmunu},
it is useful to express the background metric $g_{\mu\nu}$ in the
gauge choice \eqref{eq:ds2-N-Nphi}, so that the Hamiltonian action
reads 
\begin{eqnarray}
I\left[\Phi^{I},\Pi_{J}\right] & = & \int dx^{2}\left(\Pi_{I}\dot{\Phi}^{I}-NH-N^{\phi}J\right)\ ,\label{eq:I-Hamil-NfreeBos}
\end{eqnarray}
where $\Pi_{I}=\frac{\delta L}{\delta\dot{\Phi}^{I}}$, and $H=\frac{1}{2}\left(\Pi^{I}\Pi_{I}+\Phi^{\prime I}\Phi_{I}^{\prime}\right)$,
$J=\Pi_{I}\Phi^{I\prime}$. The deformed Hamiltonian action is then
given by 
\begin{eqnarray}
\tilde{I}\left[\Phi^{I},\Pi_{J}\right] & = & \int dx^{2}\left(\Pi_{I}\dot{\Phi}^{I}-NP-N^{\phi}J\right)\ ,\label{eq:def-freeBoN}
\end{eqnarray}
with $P=H+\sqrt{H^{2}-J^{2}}$. The transformation law of the fields
and their momenta under BMS$_{3}$ symmetries spanned by $\xi$ in
\eqref{BMS3-Vector} are then found to be 
\begin{align}
\delta_{\xi}\Phi^{I} &= \lbrace \Phi^{I},\tilde{{\cal Q}} \rbrace=\epsilon_{J}\Phi^{I\prime}+\epsilon_{P}\left(\Pi^{I}+\frac{H\Pi^{I}-J\Phi^{I\prime}}{\sqrt{H^{2}-J^{2}}}\right)\;,\label{eq:bmsTransfPhiPi} \\
\delta_{\xi}\Pi_{I} &= \lbrace \Pi_{I},\tilde{{\cal Q}} \rbrace =\left[\epsilon_{J}\Pi_{I}+\epsilon_{P}\left(\Phi_{I}^{\prime}+\frac{H\Phi_{I}^{\prime}-J\Pi_{I}}{\sqrt{H^{2}-J^{2}}}\right)\right]^{\prime}\;,\nonumber
\end{align}
where $\tilde{{\cal Q}}$ reads as in \eqref{eq:Qbms}. The field
equations $\dot{\Phi}^{I}=\{\Phi^{I},\tilde{{\cal \mathscr{H}}}\}$,
$\dot{\Pi}_{I}=\{\Pi_{I},\tilde{{\cal \mathscr{H}}}\}$, with $\tilde{{\cal \mathscr{H}}}$
given by \eqref{eq:HbmsTotal}, then follow from \eqref{eq:bmsTransfPhiPi}
by the replacement $\epsilon_{P}\rightarrow N$ and $\epsilon_{J}\rightarrow N^{\phi}$.
Note that the transformation law of $\Phi^{I}$ in the deformed theory
also reads from $\delta_{\xi}\Phi^{I}=\mathcal{L}_{\xi}\Phi^{I}$.
The transformation of supertranslation and superrotation densities
in \eqref{eq:deltaPJ-BMS3} is then recovered from those in \eqref{eq:bmsTransfPhiPi},
which goes hand in hand with the fact that $P$ and $J$ now fulfill
the BMS$_{3}$ algebra \eqref{eq:BMS3-Algebra-Cont} by virtue of
the canonical Poisson bracket $\{\Phi^{I}(\phi),\Pi_{K}(\varphi)\}=\delta_{\;K}^{I}\delta(\phi-\varphi)$.
Moreover, the stress-energy tensor of the deformed theory is obtained
from $\tilde{{\cal T}}_{\;\;\nu}^{\mu}=\sqrt{-\tilde{g}}\tilde{\Theta}_{\;\;\nu}^{\mu}$
with $\tilde{{\cal T}}_{\;\;\nu}^{\mu}$ and $\tilde{g}_{\mu\nu}$
respectively given by \eqref{eq:Tmunu-density-bms} and \eqref{eq:deformed-metric-bms}.

It is worth highlighting that the deformed action \eqref{eq:def-freeBoN}
clearly cannot be obtained from any standard limiting process of the
undeformed one for $\mathrm{N}>1$. The peculiarity of the deformed
single free boson ($\mathrm{N}=1$) stems from the fact that the supertranslation
density simplifies as $P=\Pi^{2}$, so that the momentum can be eliminated
from its own field equation, and the deformed action \eqref{eq:def-freeBoN}
can be written in Lagrangian form as 
\begin{align}
\tilde{I}\left[\Phi\right] & =\frac{1}{4}\int d^{2}x\left(\mathscr{V}^{\mu}\partial_{\mu}\Phi\right)^{2}\;,\label{eq:deform-single-Boson}
\end{align}
where $\mathscr{V}^{\mu}=\left(\sqrt{-g}\right)^{1/2}n^{\mu}$ stands
for a vector density of weight $1/2$, constructed out from the metric
$g_{\mu\nu}$ in \eqref{eq:ds2-N-Nphi} of the undeformed theory.
Noteworthy, this vector density is invariant under the BMS$_{3}$
symmetries spanned by $\xi$ in \eqref{BMS3-Vector}, since ${\cal L}_{\xi}\mathscr{V}^{\mu}=0$.
Therefore, the deformed action of a single free boson \eqref{eq:deform-single-Boson}
coincides with the ultra-relativistic limit of the undeformed theory
\eqref{eq:I-NfreeBs-CFT} for $\mathrm{N}=1$, when the Carrollian
limit is taken in a similar way as for the tensionless string \cite{Lindstrom:1990qb,Isberg:1992ia,Bagchi:2013bga,Bagchi:2015nca,Bagchi:2016yyf}.

Additionally, the vector density can be reexpressed as $\mathscr{V}^{\mu}=\frac{1}{\sqrt{2}}e^{1/2}\tau^{\mu}$,
where $e$ and $\tau^{\mu}$ correspond to the einbein and the dual
of the ``clock one-form'' of a Carrollian geometry \cite{Duval:2014uoa},
respectively; so that action of the deformed free boson agrees with
the Carrollian one found in \cite{Bergshoeff:2017btm}.

Remarkably, the action \eqref{eq:deform-single-Boson} can be understood
in terms of two inequivalent geometric structures. One of them is
Riemannian and described through the state-dependent metric $\tilde{g}_{\mu\nu}$
in \eqref{eq:deformed-metric-bms}, while the remaining structure
stands for a Carrollian manifold.

\section{Ending Remarks} 
Since the map \eqref{map BMS to CFT-1} possesses a square root, our
results also carry out for its negative branch, i.e., when the supertranslation
density is given by 
\begin{equation}
P_{(-)}=T+\bar{T}-2\sqrt{T\bar{T}}\;.\label{eq: map-negative-branch}
\end{equation}
In particular, the deformed action of a single free boson for the
negative branch reads 
\begin{eqnarray}
\tilde{I}_{\left(-\right)}\left[\Phi,\Pi\right] & = & \int dx^{2}\left(\Pi\dot{\Phi}-NP_{\left(-\right)}-N^{\phi}J\right)\ ,
\end{eqnarray}
with $P_{\left(-\right)}=\Phi^{\prime2}$. Curiously, the deformed
action $\tilde{I}_{\left(-\right)}$ agrees with an inequivalent ultra-relativistic
limit of the single free boson defined by $\Phi\rightarrow\Phi/c$,
$\Pi\rightarrow c\Pi$, when $c\rightarrow0$. This limit coincides
with that needed to pass from the standard Liouville theory to its
``flat'' version \cite{Barnich:2012aw}. Indeed, starting from a
single free boson in the conformal gauge ($N=1$, $N^{\phi}=0$) the
deformed free boson in the negative branch corresponds to the kinetic
term of the flat Liouville theory.

It is also worth to pointing out that the centrally extended conformal
algebra (given by two copies of the Virasoro algebra) can be shown
to be related to BMS$_{3}$ with central extensions, in terms of a
map that is necessarily nonlocal. Nevertheless, if only zero modes
are involved, the local nonlinear map in \eqref{map BMS to CFT-1}
still holds. Thus, blindly applying the map \eqref{map BMS to CFT-1}
for the zero modes, the Cardy formula once expressed in terms of left
and right groundstate energies (${\cal L}_{0}$, $\bar{{\cal L}}_{0}$),
given by 
\begin{equation}
S=4\pi\sqrt{-{\cal L}_{0}{\cal L}}+4\pi\sqrt{-\bar{{\cal L}}_{0}\bar{{\cal L}}}\ ,\label{eq: Cardy}
\end{equation}
reduces to its BMS$_{3}$ (or flat) version 
\begin{equation}
\tilde{S}=2\pi\frac{1}{\sqrt{-{\cal P}_{0}{\cal P}}}\left[{\cal P}{\cal J}_{0}+{\cal P}_{0}{\cal J}\right]\;,\label{eq:Cardy-flat}
\end{equation}
when the deformed energy and momentum of the groundstate ${\cal P}_{0}$
and ${\cal J}_{0}$ are expressed in terms of the BMS$_{3}$ central
charges \cite{Barnich:2012xq,Bagchi:2012xr,Bagchi:2013qva,Riegler:2014bia}.

Noteworthy, the hypotheses that ensure positivity of the Cardy formula
\eqref{eq: Cardy} (${\cal L}_{0}<0$, $\bar{{\cal L}}_{0}<0$, ${\cal L}>0$,
$\bar{{\cal L}}>0$), by virtue of both branches of the map, imply
that the deformed entropy \eqref{map BMS to CFT-1} is also positive
($\tilde{S}>0$).

Furthermore, the map between the chemical potentials follows the same
rule as that of the parameters in \eqref{map paremeters CFT to BMS},
with ($\epsilon$, $\bar{\epsilon}$)$\rightarrow$($\beta$, $\bar{\beta}$)
and ($\epsilon_{P}$, $\epsilon_{J}$)$\rightarrow$($\tilde{\beta}$,
$\tilde{\theta}$), where left and right temperatures relate to the
modular parameter of the torus as $\tau=\beta/2\pi$, and $\tilde{\beta}$,
$\tilde{\theta}$ stand for the temperature and chemical potential
of the deformed theory. Therefore, around equilibrium, the S-modular
transformation $\tau\rightarrow-1/\tau$ precisely maps into its BMS$_{3}$
(flat) version \cite{Barnich:2012xq,Bagchi:2012xr}. 
\begin{equation}
\tilde{\beta}\rightarrow\frac{4\pi^{2}\tilde{\beta}}{\tilde{\theta}^{2}}\;\;,\;\;\tilde{\theta}\rightarrow-\frac{4\pi^{2}}{\tilde{\theta}}\ .
\end{equation}

As a closing remark, since the uplift of the deformed action \eqref{eq:I-tilde-detTmunu}
to higher dimensions is clearly invariant under diffeomorphisms and
local scalings, it would be worth exploring whether the $D$-dimensional
deformed theories might be invariant under the conformal Carrollian
algebra, which is known to be isomorphic to BMS$_{D+1}$ \cite{Duval:2014uva}.

\medskip{}

\section*{Acknowledgments}
We thank Glenn Barnich, Diego Correa, Gast\'{o}n Giribet, Joaquim Gomis,
Hern\'{a}n Gonz\'{a}lez, Fabio Novaes, Carlos Nunez, Pulastya Parekh, Alfredo
P\'{e}rez, Miguel Riquelme and Guillermo Silva for useful comments and
discussions. This research has been partially supported by ANID FONDECYT
grants N$^{\circ}$ 1171162, 1181031, 1181496, 1211226.

\end{document}